\documentstyle[aps,pre,preprint]{revtex}
\newcommand{\bap}{\bar{P}}
\newcommand{\baq}{\bar{Q}}
\begin{document}
\title{\bf Non equilibrium mode coupling theory for supercooled
liquids and glasses}

\author{A. Latz}

\address{Institut f{\"u}r Physik, Johannes Gutenberg Universit{\"at}, 
Staudinger Weg 7, D-55099 Mainz, Germany}

\date{\today} \maketitle
\begin{abstract}
A formally  exact set of equations is derived for the description of
nonequilibrium phenomena in classical liquids and glasses.  With the
help of a non equilibrium projection  operator formalism, the correlation
functions and fluctuation propagators are expressed in terms of
memory functions  and time dependent collective frequencies.   This
formally exact set of equations is approximated by applying
mode coupling approximations to the memory functions. The resulting set
of equations for wavevector dependent correlation functions,
fluctuation propagators  and one-time structure factors $S_q(t)$
generalizes the well known mode coupling theory of the glass
transition to situations far away from equilibrium. 
\end{abstract}

\section{Introduction}
\label{sec:introduction}
The mode coupling theory of the glass transition (MCT) is based on the
experimental and theoretical understanding of dense liquids.  Although 
approximative in nature, 
mode coupling theories were quite successful to describe
quantitatively 
the dynamics of dense liquids close to the triple point. 
The dynamics of dense liquids is dominated by two mechanisms, the
precursor of the cage effect and a complicated coupling of current
fluctuations and density fluctuations. This coupling is the main cause of
activated motion close to the glass transition \cite{hopping,dasmaz} 
and will not be considered 
further in this paper. The cage effect is a collective
phenomenon caused by the mutual hindering of the motion of the atoms or
molecules due to the  strong repulsion of the cores of the atoms.  Instead of
a unidirected flow-like motion a kind of stop and go mechanism is observed in
simulations of dense liquids. The particles tend to rattle in  a cage formed
by the  surrounding atoms. To escape this cage an atom has to find a
hole in the
surrounding wall formed by the other atoms. 
The same is of course true for the atoms
building the wall. Due to the thermic motion a situation will arise in which
the particles can pass each other. This process will be repeated over and over
again. In this way the motion of particles, which are
nearby at some instant of time, will decorrelate slowly. In the
process a
back flow pattern is created.   
By increasing the density, the local cage effect will
eventually lead to a complete blocking of long range motion of the
particles. They will be stuck in their local cages and thus form a
glass. The details how 
this is done, especially how this glassification affects the various
spatial scales are described by the idealized mode coupling theory
\cite{BGS}.
Within this theory it is e.g. possible to understand the effect of
the stiffening of 
the local cages on the thermodynamics and hydrodynamics close to the glass 
transition \cite{goela}. 
Experimentally observed phenomena like the frequency dependent specific heat 
are thus explained quite naturally in a unified framework.

The underlying mathematical mechanism of the glass transition in the idealized
MCT is a bifurcation in the equation for the dynamics of the density - density
correlation function of a system of N particles.  Using a standard Mori 
projection operator formalism for
the density and current fluctuation  $n_q(t), \vec{j}_q(t)$ with 
\begin{eqnarray}
n_q(t) = \exp(i L t) n_q(0) = \sum_i^N \exp(i \vec{q} \vec{r}_i
(t))\label{4}\\ 
j_q(t) = \frac{\vec{q}}{q} \vec{j}_q(t) = \sum_i^N \frac{\vec{q}}{q} \vec{v}_i 
\exp(i \vec{q} \vec{r}_i(t)) \label{5},  
\end{eqnarray}
the following formally exact
equation for the density
density correlation function $\phi_q(t)$ can be derived:  
\begin{eqnarray} 
\frac{d \phi(t)}{d t} &=& i \vec{q} \phi_{n,\vec{j}}(t)  :=
i q \phi_{n,j}
\label{1} \\
\frac{d^2 \phi(t)}{dt^2} &=& - \Omega_q^2 \phi(t) - \int_0^t d t' \frac{d
  \phi(t')}{d 
  t'} M(t-t'). \label{mct}  
\end{eqnarray}

Here the correlation function $\phi$ is given  as
\begin{equation} \label{3}
\phi(t) \equiv \phi_{nn}(t) = (n_q|n_q(t)):=  Tr(\exp(- \beta H) n^*_q(0)
n_q(t)) 
\end{equation}
$L$ is the Liouville operator. Its action on  some dynamic variable is defined
by the Poisson bracket 
\begin{equation}\label{6} 
L A = -i \sum_i^N \{H, A\}.
\end{equation}
By construction the memory function $M$ is free of any 
hydrodynamic singularity 
and cannot be expressed as a linear functional of the 
correlation functions. Instead it
can  be written as a correlation function of fluctuation forces.
\begin{equation} \label{7}
M_q(t) = (Q L j_q| \exp(i L Q t) Q L j_q)/S_{jj}. 
\end{equation} 
$S_{jj}$ is the static current current fluctuation function. In equilibrium it
is given as 
\begin{equation}\label{sjj}
S_{jj} = N k_B T /m
\end{equation}
where $m$ is the mass of an atom and $T$ is the equilibrium temperature. 
In the mode coupling approximation the fluctuating forces are approximated as
a superposition of two mode contributions \cite{BGS} 
\begin{eqnarray} 
Q L j_q &\approx& \sum_{k p} v(q|kp)n_k n_p \label{8}\\
v(q|kp) &=&  (Q L j_q| n_k n_p) (S_k S_p)^{-1} =
\delta_{\vec{p},\vec{q}-\vec{k}} S_{jj} \hat{v}(k,q-k) \label{9} 
\end{eqnarray}   
The second equality in eq. (\ref{9}) is due to momentum conservation. $S_q$ is
the static structure factor $\phi(q,t=0)$. The vertex $\hat{v}$ can be
evaluated in equilibrium. Neglecting static direct three point correlation 
functions the memory function is given by \cite{BGS} 
\begin{equation}\label{mem} 
M_q(t) = n (k_B T/m)^2 \frac{1}{2}
 \sum_k  \hat{v}^2(k,q-k)  \phi(q-k,t) \phi(k,t)
\end{equation} 
with 
\begin{equation}
\hat{v}(k,q-k) = ( \vec{k} \frac{\vec{q}}{q} c_k + (\vec{q}-\vec{k}) 
\frac{\vec{q}}{q} c_{q-k})^2 \label{vertstat}
\end{equation}
Here $c_q$ is the (equilibrium) direct correlation function. 
Eqn \ref{mct} and \ref{mem} form
a closed set of equations for the density -  density correlation
function in equilibrium. The solution properties of the above
equilibrium MCT (EMCT) equation are very
well known \cite{review-glass}.  They exhibit an ideal glass transition
as a bifurcation scenario from an ergodic to a non ergodic state. The
transition temperature $T_c$ (or density, pressure ...) is independent of
the cooling rate since it is an equilibrium property of the system. By
approaching $T_c$ the structural relaxation is dramatically slowing
down, developing the typical two step relaxation pattern of $\beta$ - 
and $\alpha$ - relaxation.  
The EMCT was quantitatively tested for a experimental realization of a
hard sphere system \cite{colloid}.  Without any
fit factor  (up to an overall
scale factor) it is possible to reproduce with 20 $\%$ accuracy  
the density - density correlation for the wavevector at the peak
of the structure factor.  
With the same kind of approximations for
memoryfunctions the theory was also quantitatively tested for a binary 
Lennard Jones systems \cite{nauroth}, a system of dumbbells
\cite{kaemmerer} and a model system for water \cite{water}. 
Any generalization to non equilibrium situations should therefore in
equilibrium  reduce to the above set of equations (\ref{mct}) and (\ref{mem}). 

The non equilibrium situations to be described in this paper arise if
one investigates the 
relaxation in a system, which is not yet
equilibrated. When cooling down from some temperature
$T_1> T_c$ to $T_2 > T_c$, one has to wait at least for some time $t_w
$ comparable to a typical $\alpha$ - relaxation time $\tau_q(T_2)$ to
guarantee equilibration at the temperature $T_2$. As indicated,
the $\alpha$ - relaxation time $\tau_q$ depends on the wavevector $q$.
To use the EMCT, density fluctuations for all wavevectors
have to be equilibrated. For hard sphere like systems the largest
$\alpha$ - 
relaxation time is given by the wavevector at the peak of the
structure factor i.e only for a time $t_w > 
\tau(q_{max})$ the above equations are able to describe the density 
correlation functions. For times smaller than $\tau(q_{max})$
nonequilibrium effects like breaking of time translation invariance
are to be expected. E.g. the one-time structurefactor $S_q(t) =
\phi_q(t,t)$ is time dependent, since the probability density has not
yet reached its equilibrium value.  

As soon as the system is quenched below
$T_c$, the behavior of the system changes qualitatively. 
In EMCT the $\alpha$ relaxation time is going to infinity for all
wavevectors, when $T$ is approaching $T_c$ from above, i.e. 
the system never reaches equilibrium for $T\le T_c$. 
This means that e.g. even for infinite long waiting time the one-time 
structure factor will be different
from the value, it would reach in the presence of hopping processes.   
For real systems, where activated processes cannot be neglected, the
waiting time has to be smaller than the inverse of the typical
hopping rate, in order to see aging. 
Therefore the applicability of EMCT has to be
reconsidered within the larger framework of a theory, which is
applicable also in nonequilibrium situations. 

In the theory of spin glasses quite some progress in this direction was
obtained in recent years. Already Kirkpatrick, Thirumalai
  \cite{ted} showed that the
dynamics  of an infinite ranged p - spin system with $p \ge
2$  and a disordered Potts system ($p > 4$) exhibit
great similarities to the dynamics of structural glasses at least
close to the critical temperature $T_c$ (which is in the spin glass 
literature called $T_D$).  The equation for the correlation function
for $T> T_c$ reduce to a schematic EMCT model of the glass
transition. For $p=3$ the p -  spin model is equivalent to the 
so called Leutheusser model \cite{Leutheusser} above $T_c$. 
This model exhibits a glass transition, but its  
$\alpha$ - relaxation is a simple exponential relaxation and can 
therefore not explain the typical stretched relaxation observed in
real glasses. \footnote{It is possible to consider a more general class of
schematic models, which show non exponential relaxation
\cite{review-glass}, but they are not easy to justify within the
theory of liquids.} In the context of superccoled liquids the
Leutheusser model can only be motivated by
the complete neglect of the wave vector dependence in
equ. (\ref{mct},\ref{mem}).  For the p - spin model the
Leutheusser model is an exact theory due to the
mean field character of this system.  Cugliandolo and
Kurchan showed for the p - spin model , that the  EMCT is not valid
any more below $T_c$, due to similar reasons
as explained above \cite{ck}.  Instead they derived nonequilibrium 
mode coupling equations for this
model and solved them asymptotically below $T_c$.  Contrary to the
EMCT, the dynamics cannot be understood in terms of correlation
function alone, but a coupled set of equations for the correlation
functions $\phi(t_w,t)$ and susceptibilities
$\chi(t_w,t)$ is necessary. 

The most interesting property of the solution of these equations is the
separation of the dynamics in a short time part, where the fluctuation
dissipation theorem (FDT) is obeyed and an aging part where FDT is
violated in a very specific way. In the aging regime the equations for 
$\phi(t_w,t), \chi(t_w,t)$  can be reduced to a single equation for a
function $\hat{\phi} (\lambda)$ with $\lambda = h(t_w)/h(t)$. The
susceptibility is given as 
\begin{eqnarray}
\chi(t_w,t) &=& \frac{h'(t_w)}{h(t)} \hat{\chi}(\lambda) \label{scaling}\\ 
\hat{\chi}(\lambda)& = & \frac{x}{T} \frac{d \hat{\phi}}{d \lambda}
\nonumber
\end{eqnarray} 

The function $h$ has to be determined by matching the FDT solution and
the aging regime. So far its form is only known for the spin glass with
$p=2$ \cite{SK}.  The  term $\frac{T}{x}$ was interpreted as
an effective temperature \cite{fictive}. 
Since the important work of Cugliandolo and Kurchan
a large number of papers has addressed the problem of nonequilibrium
relaxations in mean field models \cite{aginglist}. It was shown for 
general nonlinear $\Phi^n$ fieldtheories that
in the cases where  NMCT equations  
are exact, a mapping to an equivalent disordered
problem can be found \cite{boucug}.  This can also be achieved by
deterministic but highly irregular interactions \cite{franz}.  

Simulation of binary liquids in the out of equilibrium regime 
\cite{parisi_bin,bako1,bako2,bako3} 
exhibit striking qualitative similarities to the dynamics found in the p -spin
model. There is the separation in FDT and aging regime. Also the
scaling behaviour (\ref{scaling}) seems to be fulfilled.
It is worth mentioning, that the crossover from the FDT regime to
the aging regime, i.e. the behaviour of the function $x(t_w,t)$, seems to
be different than in the $p$ -spin model.  To be able to describe
behaviour of real liquids the schematic  models are not sufficient,
since they lack the important wavevector dependence, leading e.g. to
non exponential relaxation and experimentally testable results like 
wavevector dependent Kohlrauschexponents \cite{fula}.  Equations for
multicomponent models have been formulated on an abstract level
\cite{boucug,manifold}, but there relation to the theory of glass
forming liquids is not obvious. 
Additionally there are no equations for
the one time structure factor $S_q(t)$ for real liquids.  In the
following, I am going to formulate a generalization of the EMCT for real
liquids, which will enable the study of the dynamics of real liquids 
above and below $T_c$ in equilibrium and nonequilibrium situations.

\section{The nonequilibrium projection operator formalism}
The main reason for the appearance of generic nonequilibrium
relaxation is the inability of the system to equilibrate below $T_c$
(if hopping is neglected, which I will assume in the following). To keep
the formalism as simple as possible, I will consider a situation where
the probability density $\rho_{NE}$ is different from the canonical ($\rho
\ne \exp (- \beta H)$)  or other
equilibrium densities but the microscopic dynamics is still
Hamiltonian. The only requirement for not being in equilibrium  is 
\begin{equation}
 \{\rho_{NE}, H\} \ne 0
\end{equation}

This situation may e.g. arise if the system is isolated after an
initial quench from an equilibrated liquid above $T_c$ to a glass
below $T_c$ (e.g. by coupling to an external heat bath, whose
temperature is below $T_c$ ). The system tries to get into equilibrium
with the heat bath. As argued in \cite{goela} it can be expected that the
kinetic energy is equilibrating on microscopic time scales. It is the
potential energy which does not reach its equilibrium value due to
its contribution from structural relaxations. After the initial
quench we may therefore expect to describe the system with an initial
density operator $\rho_{NE} = \exp(-(\beta_{f} K + \beta_I
\epsilon^{pot})$. Here $\beta_{I}$ is the inverse of the initial
temperature before the quench and $\beta_{f}$ is the inverse of the
final temperature. For $\beta_c = 1/(k_B T_c)$ the relation 
$\beta_I < \beta_c < \beta_{f}$   is fulfilled.  $K,
\epsilon^{pot}$ is 
the kinetic and potential energy respectively.
If needed the formalism can also be developed by assuming a
coupling to an external heat bath by using a Gaussian thermostat to
guarantee the average of the kinetic energy being constant at any
time \cite{me}, but here I want to concentrate on the more simple
case of an isolated system described by a nonequilibrium density operator.   
 
As basic variables for the following projection operator formalism
(POF)  $A_i(t)$ I again will use density fluctuations (\ref{4}) and
current fluctuations (\ref{5}). 
The energy fluctuations are neglected. They are of course
influenced by the glass transition, but they do not qualitatively
change the glass transition scenario \cite{goela}.  The main idea of
any projection operator formalism  is to  
separate the dynamics of the variables in a part
directly proportional to a chosen variable, a part which is given by
the fluctuation of the variables in the past  and a fluctuating force,
whose dynamic was never directly proportional to the fluctuations of
the basic variable $\delta A_i(t)$.  
The most direct generalization of the standard Mori formalism to
situations, in
which also equal time correlation functions are time dependent is
obtained by using \cite{kawagunton,grabert,furukawa} 
\begin{eqnarray}
\bap(t) &=& \exp(-i L t) \delta A_i(0) {\bf S}^{-1}_{ik}(t)\delta
A_k(t) \exp(i L t) \label{p}  \\
&:=& \exp(-i L t) |A_i(0)) {\bf S}^{-1}_{ik}(t) (A_k(t)|
\exp(i L t) \nonumber \\    
\baq(t) &=& 1 - \bap(t) \label{q}
\end{eqnarray} 
Equal indices are summed over. In equilibrium 
this projection operators reduces to the time independent Mori projection
operator in equilibrium. The matrix $\bf S$ is the equal time
correlation matrix of the variables $A_i$ 
\begin{equation}\label{Sij}
{\bf S}_{ik} = (A_i(t)|A_k(t)) = Tr(\rho_{NE}\; exp(i L t) A_i(0)
|\exp(i L t)A_k(0)). 
\end{equation}
The projection operators  have the following properties. 
\begin{eqnarray}
\bap(t_1) \bap(t_2) = \bap(t_2) \nonumber\\
\baq(t_1) \baq(t_2) = \baq(t_1) \label{proprop}\\
\baq(t_2) \bap(t_1) = 0. \nonumber
\end{eqnarray}

With the help of (\ref{p},\ref{q}) a formally exact set of
equations for the fluctuation of the density (\ref{4}) and the current
(\ref{5}) can be derived
 
\begin{eqnarray}
\lefteqn{\frac{d n_q(t)}{d t} =   i q j_q(t)} \label{cont}\\
\lefteqn{\frac{d j_q(t)}{d t} = i L j_q = } \label{strom}\\
&&i \delta n_q(t) q^2 \Omega_{nj}(t) + i \delta j_q(t)
q^2 \Omega_{jj}(t) \nonumber \\
&&  - \int_{t_w}^t dt' \delta n_q(t') {\bf S}^{-1}_{nj}(t') 
M_q(t',t)  
 - \int_{t_w}^t dt' \delta j_q(t') {\bf S}^{-1}_{jj}(t') 
M_q(t',t)  \nonumber \\
&& + i \exp(i L t_w) \baq(t_w) G(t_w,t) L \delta j_q(0) \nonumber
\end{eqnarray}

The structure of the equation is very similar to the equilibrium
equation. Eq. (\ref{cont}) is simply the continuity equation for the
density. $ q^2 \Omega_{nj},q^2 \Omega_{jj}$ are generalized time dependent
frequencies. They are given by 
\begin{eqnarray}
\Omega_{nj} &=& S_{nn}^{-1}(t) (n_q(t)| L j_q(t))/q   + S_{nj}^{-1}(t)
(j_q(t)| L j_q(t))/q \label{freq}\\
\Omega_{jj} &=& S_{jn}^{-1}(t) (n_q(t)| L j_q(t))/q   + S_{jj}^{-1}(t)
(j_q(t)| L j_q(t))/q \label{freqjj}
\end{eqnarray}

In equilibrium $\Omega_{jj}$ would be zero due to time translational
invariance, but has to be kept in nonequilibrium for finite times.
The memory function $M_{q}$ is an explicit function of two times. 
\begin{eqnarray}
M_q(t_1,t) &=& (L j_q(t_1)| \exp(i L  t_w) \baq(t_w) G(t_w,t)
L \delta j_q(0)) 
\label{memne}\\
&&G(t_1,t) = {\cal T} \exp ( i \int_{t_1}^t d t' \, L \baq(t'))
\label{g}
\end{eqnarray}
Where the operator ${\cal T}$ induces a time ordering from left to
right. It is important to note that in general the memory function
(\ref{memne}) does not have   
the form of a correlation function of fluctuating forces, but is the
correlation of the accumulated random force between $t_1$ and $t$ and
the real force $L j_q(t_1)$ at time $t_1$. It describes the
dissipation between $t_1$ and $t$ \cite{grabert}. This will be
important for finding appropriate  approximations. 
The last term in (\ref{strom}) is a fluctuating force $F_q(t_w,t)$
\begin{equation} \label{random}
F_q(t_w,t) = \exp(i L t_w) \baq(t_w) G(t_w,t) L j_q(0).
\end{equation}
Due to the properties of the projection operators (\ref{proprop}) the time
ordered product in (\ref{g})   propagates the force $L j_q$ always
in the space of functions perpendicular to density and current at every
time step. The fluctuating force $F_q(t_w,t)$ is not correlated with the
fluctuation of density or current at time $t_w$.   

Equ. (\ref{cont},
\ref{strom}) have the form of an inhomogeneous integro differential
equation. They can  be solved formally by
introducing fluctuation propagators $N_{kl}$, which obey the
homogeneous part of equ. (\ref{cont},
\ref{strom}). 

\begin{eqnarray} 
\frac{d N_{in}(q;t_1,t)}{d t} &=&   i q N_{ij}(q;t_1,t) 
\label{dichtprop}\\
\frac{d N_{ij}(q;t_1,t)}{d t} &=&i  N_{in}(q;t_1,)
\Omega_{nj}(t) + i  N_{ij}(q;t_1,t) 
\Omega_{jj}(t)  \label{stromprop}\\
&&  - \int_{t_w}^t dt' N_{in}(q;t_1,t') {\bf S}^{-1}_{nj}(t') 
M_q(t',t) \nonumber \\ 
&& -\int_{t_w}^t dt' N_{ij}(q;t_1,t') {\bf S}^{-1}_{jj}(t') 
M_q(t',t)  \nonumber 
\end{eqnarray}
The initial conditions are $N_{ik}(q;t,t) = \delta_{ik}$.
If a density or current fluctuation is given at a time $t_1$ e.g. by
applying a known external field to the system, the fluctuations at all
later times are then given given by 
\begin{eqnarray} 
\delta n_q(t) &=& \delta A_i(t_1) N_{in}(q;t_1,t) + i \int_{t_1}^t
F_q(t_1, t') N_{jn}(t',t) \label{flucdicht}\\
\delta j_q(t) &=& \delta A_i(t_1) N_{ij}(q;t_1,t) + i \int_{t_1}^t
F_q(t_1, t') N_{jj}(t',t) \label{flucstrom},
\end{eqnarray}
where $F_q(t_1,t)$ is the random force eq. (\ref{random}). The $A_i$
are density and current respectively for $i =1,2$. This can easily be
checked by replacing density and current in equ. (\ref{cont},\ref{strom})
by (\ref{flucdicht}, \ref{flucstrom}).

The fluctuation propagators are very closely related to the 
susceptibilities. These are given by Poisson brackets of the variables
\cite{grabert}. 
\begin{equation} \label{suscept}
\chi_{ik} = \langle \{\delta A_i^*(t_1),\delta A_k(t) \}\rangle 
\end{equation}

The density susceptibility can therefore be written with the help of
eq. (\ref{flucdicht}) as 
\begin{equation}\label{suscep2}
\chi_{nn}(t_1,t) = -i q N_{jn}(t_1,t) + i \int_{t_1}^t
\langle \{n^*_q(t_1),F_q(t_1,t')\} N_{jn}(t',t) \rangle
\end{equation}

Here it was used that $\chi_{nn}(t_1,t_1) = 0$ and
$\chi_{nj}(q;t_1,t_1) = -i q$. In mode coupling approximation,
the second term in  (\ref{suscep2}) is zero \cite{me2}.
The dynamical susceptibility can therefore be identified with the
fluctuation propagator $-i q N_{jn}(q;t_1,t)$. 

The equation for the density correlation function can be easily derived from
eqs. (\ref{cont},{\ref{strom},\ref{flucdicht}).  

\begin{eqnarray}
\frac{d^2 \phi_q(t_1,t)}{d t^2} &=&  
-\phi_q(t_1,t) \; q^2 \Omega_{nj}(t) + i \frac{d \phi_q(t_1,t)}{dt}
 q \Omega_{jj}(t) \nonumber  \\
&&  - i q \int_{0}^t dt' \phi_q(t_1,t') {\bf S}^{-1}_{nj}(t') 
M_q(t',t)\label{correl} \\ 
&& - \int_{0}^t dt' \frac{\phi_q(t_1,t')}{d t'} {\bf S}^{-1}_{jj}(t') 
M_q(t',t)  \nonumber \\
&& + \int_0^{t_1} (-i q N_{jn}(q;t',t_1)) \bar{M}_q(t',t) \nonumber
\end{eqnarray}

The function $\bar{M}_q(t',t)$ is in general not identical to the
memory function $M_q(t_1,t)$ in (\ref{memne}), but it is the
correlation function of the fluctuating forces.
\begin{equation}\label{corrfluc}
\bar{M}_q(t_1,t)= (\baq(0) F_q(0,t_1)|\baq(0) F_q(0,t))
\end{equation}

Eqs. (\ref{correl},\ref{stromprop}) are  a formally exact set of
equations for density correlation function and fluctuation propagators,
which reduces to the response or integrated response within mode
coupling theory. 
If we assume equilibrium, i.e. if we use the canonical ensemble, the set of 
equations reduces exactly  to the one in equilibrium \cite{me2}. 
Time translation invariance is restored, the fluctuation propagator 
$N_{nn}(t_1,t)$ is identical to the normalized
correlation function, $\bar{M} = M$ and the equation (\ref{correl})
reduces to (\ref{mct}) with $q^2 \Omega_{nj} = \Omega_q^2$.  

For nonequilibrium situations a further step has to be
taken. In EMCT the only input needed is the static structure
factor. The frequency $\Omega_q$ can easily be calculated if $S(q)$ is
known. In nonequilibrium $S(q)$ turns into a time dependent
function. E.g. for quenches we only know the value of $S_q(t=0)$,
which is just the equilibrium structure factor of the liquid at the
begin of the quench. The equations presented so far, do not provide an
equation for the equal time correlation function. It is easily
possible to obtain an equation by differentiating twice $S_q(t)$. 

\begin{equation}\label{struct1}
\frac{d^2 S_q(t)}{d t^2} = 2 q^2 (j_q(t)|j_q(t)) - 2 Re((n_q(t)|L j(t)) 
\end{equation}

Equations like this cannot be treated with the standard non
 equilibrium projection operator formalism, since it is assuming
 implicitly that frequencies like $ q (n_q|L j_q)$ are known or can
 be simply calculated. This is not the case for liquids close to the
 glass transition in nonequilibrium. Due to $L \rho_{NE} \ne 0$ the
 Liouville operator is not a self adjoint operator like in
 equilibrium. Therefore $(n_q|L j_q) \ne (L n_q|j_q)$ and can therefore
 not simply be calculated. One possibility is to express the frequencies  
$\Omega_{nj}(t)$ and $\Omega_{jj}(t)$ with the help of a modified projection
operator \cite{me2} by the one time structure factor, correlation functions
and susceptibilities. The equation (\ref{struct1}) can therefore be written in
 the form 
\begin{eqnarray} 
\frac{1}{2} \frac{d^2 S_q(t)}{d t^2} &=&  
S_{jj}(t) - Re[S_q(t) \; q^2 \Omega_{nj}(0) + S_{nj}(q;t) 
q^2 \Omega_{jj}(0) \nonumber  \\
&&  + i q \int_{0}^t dt' \phi_{nl}(t,t') {\bf S}^{-1}_{lm}(0) 
W_{mj}(q;t',t) \label{struct2} \\ 
&& -q \int_{0}^t dt' \frac{\phi_{nl}(q;t,t')}{d t'} {\bf S}^{-1}_{lm}(0) 
U_{mj}(t',t)  \nonumber \\
&& - \int_0^{t} (-i q N_{jn}(q;t',t))^* \bar{M}_q(t',t)] \nonumber
\end{eqnarray}

There appear 4 new memory functions $W_{mj}$ and $U_{mj}$ with $m=1,2$
in this equation. $W_{mj}$,
$U_{mj}$ are the
correlation functions of the fluctuation of the variable $A_m(0) \in (n_q,j_q)$
and the current of the fluctuating force $L F_q(t',t)$ between $t'$ and $t$
or the fluctuating force respectively, translated backwards in time to $t=0$.  
\begin{eqnarray} 
W_{mj}(t',t) &=& (A_m(0)|\exp (- i L t') L F_q(t',t)) \label{W}\\
U_{mj}(t',t) &=& (A_m(0)|\exp (- i L t') F_q(t',t)) \label{U}
\end{eqnarray}

In equilibrium the memory function $U$ vanishes exactly. The fluctuating
force $F_q(t',t)$ is perpendicular to $A_m(t')$ and  $U_{mj}(t',t) \equiv 
(\exp (+ i L t')A_m(0)| F_q(t',t)) \equiv 0$. The memoryfunction $W_{nj}$
is in equilibrium equal to the fluctuating force correlation function
$\bar{M_q}$. Together with the fluctuation dissipation theorem therefore the
convolution integrals in time in (\ref{struct2}) cancel. The first line in
(\ref{struct2}) vanishes, since in equilibrium  $\Omega_{jj} 
\equiv 0$ and $q^2 \Omega_{nj} = S_{jj}/S_q$.  
Equations (\ref{struct2},\ref{correl},\ref{dichtprop}) and (\ref{stromprop})
are an exact set of equations for one time and two time quantities,
describing the change of the structure and the correlations and susceptibilities
if e.g. by an initial quench a nonequilibrium probability density is imposed.

\section{Approximations}
To close the equations, closure relations for the memoryfunctions $M_q,
W_{mj}, U_{mj}$  and
the correlation function of fluctuation forces $\bar{M}_q$ have to be found.
There is no systematic way of calculating the memory function, since no small
parameter is available for dense liquids close to the glass transition. The
only guide available is the requirement, that the equations should reduce to
the equ. (\ref{mem}) and (\ref{mct}) in equilibrium and that certain
mathematical requirement intrinsic to the projection operator formalism are
fulfilled. It is very plausible to
express $\bar{M}$ as a quadratic form of the correlation functions
since it has the 
structure of a correlation function of fluctuating forces. 
\begin{equation}
\bar{M}(q;t_1,t) = \frac{1}{2} S_{jj}^2(t) \sum_{k}  v^2(q;k,q-k;t)
\phi_k(t_1,t) 
\phi_{q-k}(t_1,t) \label{approm}. 
\end{equation}  
The vertex $v(q;k,q-k;t)$ is the same as in (\ref{vertstat}) if
the direct correlation functions are understood as time dependent
quantities. 
The memoryfunction $M$ describes the influence of a force fluctuation at
$t'$ on its surrounding integrated between $t'$ and $t$. Since $M$
cannot be written as a linear functional of the integrated response
the next simple approximation possible is the following:  
\begin{eqnarray}
\lefteqn{M(q;t_1,t) = M_q(t,t) -}\label{appromb} \\
\; \; S_{jj}^2(t_1)\sum_{k} \int_{t_1}^t d t' v^2(q;k,q-k;t')&& \nonumber
(-i k) N_{jn}(k;t',t)
\phi_{q-k}(t',t) 
\end{eqnarray}
where $N_{jn}(k;t',t)$ can be identified with the susceptibility as
explained above. In equilibrium this guarantees , that $M_q =
\bar{M}_q$. Therefore the two last integrals in eq. (\ref{correl})
cancel for times
$t_1 = t$. This property has to be fulfilled 
rigorously \cite{me}. It can therefore be used to calculate $M(t,t)$
selfconsistently. Although $M$ and $\bar{M}$ are equal in equilibrium,
the formal expressions in nonequilibrium are different. Choosing $M = \bar{M}$
has the consequence that a generalized fluctuation theorem of the second kind
\cite{grabert} would be fulfilled. It can be shown \cite{me} that this 
immediately excludes the possibility
of the aging properties found e.g. in simulations \cite{bako1}
-\cite{bako3}. Only a transient aging phenomena caused by the time dependence
of the static structure factors would be observed in this case.   
Another possibility to introduce approximations is first to 
integrate the third integral in (\ref{correl}) by parts. This leads to 

\begin{eqnarray}
\frac{d^2 \phi_q(t_1,t)}{d t^2} &=&  
-\phi_q(t_1,t) \; q^2 \Omega_{nj}(t) + i \frac{d \phi_q(t_1,t)}{dt}
q \Omega_{jj}(t) \label{correl2} \\
&& - S_q(t) M(t,t) + \phi_q(t,0) M(0,t) \nonumber \\
&& + \int_{0}^t dt' \phi_q(t_1,t')\Sigma(t',t) \\ 
&& + \int_0^{t_1} (-i q N_{jn}(q;t',t_1)) \bar{M}_q(t',t) \nonumber
\end{eqnarray}

with
\begin{equation} \label{sigma}
\Sigma(t',t) = 
 {\bf S}^{-1}_{jj}(t') 
\frac{d M_q(t',t)}{d t'} - i q {\bf S}^{-1}_{nj}(t') 
M_q(t',t)
\end{equation}

With the approximation 
\begin{equation} \label{approxck}
M(q;t_1,t) = S_{jj}^2(t_1) \sum_{k}  v^2(q;k,q-k;t) N_{jn}(k;t_1,t)
\phi(q-k;t_1,t) 
\end{equation}
the resulting structure of the theory is very similar to the 
one of the p - spin model \cite{ck}.
This is also true
for the theory introduced by the approximations (\ref{approm}, \ref{appromb}).
Both reduce  to EMCT
in equilibrium. Which of the
approximations is more appropriate for real liquids has to be tested
and is part of future work.

Similar approximations are possible for the new memory functions $U$ and
$W$. But since there are less mathematical boundary conditions to be
fulfilled than for $M$ and $\bar{M}$ there is more freedom in the choice 
of the approximation. Different approximation will lead only to 
quantitatively but not qualitatively different results for $S_q$. 

\section{Conclusions}
 
With the help of non equilibrium projection operator techniques it is possible
to derive an exact set of equations for one-time quantities like the time
dependent equal time structure factor and two-time quantities like correlation
functions and fluctuation propagators. The memory functions can be approximated
as in the EMCT by the correlation functions and response functions. 
It is interesting to note that in the formulation
presented here, a theory for the equal time structure factor (which
is trivial in the p - spin model) and not the
potential energy as in the p -spin model is derived. The structure factor is
directly accessible to experiments and it would be especially interesting to
study aging phenomena in colloidal hard sphere glasses as it is done
for colloidal clays \cite{bonn},
where the structure factor can
easily be measured within light scattering experiments. It is already obvious
from the equations, that the structure factor $lim_{t \to \infty} S_q(t)$ in
the glass will not be the structure factor of the corresponding equilibrium
liquid, but will depend on the nonergodicity parameters i.e. on $\lim_{t \to
  \infty} \lim_{t_w \to \infty} \phi_q(t_w,t)$ via the memory functions $
W_{jj}, U_{nj}$ and the correlation function of fluctuating forces $\bar{M}$. 
The aging scenario to be expected from the equations presented here can be 
formulated within the bifurcation scenario of the glass transition derived
from mode coupling theory
\cite{review-glass}.  This picture can best be explained in discretized models,
where only a discretized set of wave vectors is used. In this case the
wavevector components of the equilibrium structure factor $S^{equ}_q$
can be considered as control parameter 
of the mode coupling models. In the space formed by the components exists a 
critical hypersurface which separates the control parameter space in a liquid
subset and a glassy subset. Since in the nonequilibrium case  the
value of $S_q(t)$ for $t \to \infty$ depends on the nonergodicity
parameters, the 
critical hupersurface will also depend on the two time correlation function. A
quench can now be considered in the following way. By starting in the liquid
subset the system, which can be considered as a a point in the control
parameter space, starts to 
flow towards the critical hypersurface. As it approaches this 
surface relaxations become slower and slower.  
Due to the glass transition at the
critical hypersurface the system can not reach the point, which would
correspond to the thermodynamic parameters in equilibrium. Instead  it will 
very likely be trapped on the (nonequilibrium) critical hypersurface, which 
also depends on the nonergodicity parameters. In hard core like systems the
glass transition is driven by the nearest neighbour repulsion. This can be
inferred from the fact, that in the EMCT
equations the glass transition occurs, if the first peak of the structure
factor becomes big enough. Therefore trapping on the critical hypersurface
will lead to a drastic change in the sensitivity of the peak height on the
quench temperature (or density), once the system is quenched below $T_c$ (above
$n_c$).  This qualitative aspect can be seen in simulations of binary Lennard
Jones sytems 
\cite{bako3}. In this simulation the area under the first peak of the 
radial distribution
function $g(r)$, which roughly corresponds to the height of the first peak of 
the structure factor $S_q$, does not seem to increase beyond a certain
threshold, if the system is quenched below $T_c$. 
As a next step the details of the theory presented will be
studied in simplified models before calculating the properties of  
realistic glass forming liquids. This will be part of future studies. 
\vspace{1cm}

\noindent {\bf Acknowledgments}

\noindent I would like to thank R. Schilling for his interest and support
during the development of the formalism and Leticia Cugliandolo and
Walter Kob for 
interesting discussions and a critical reading of the manuscript. 
The work is partially supported by the Sonderforschungsbereich 262 of
the Deutsche Forschungsgemeinschaft.


\end{document}